\documentclass[epj,referee]{svjour}
\usepackage{graphics}

\begin{document}

\title{Energy and entropy effects of counterions in salt-free colloidal solutions}

\author{Chi-Lun Lee\inst{1}}

\institute{
  \inst{1} Department of Physics, National Central University, Jhongli 32001, Taiwan\\
}

\date{\today}

\abstract{
We use a shell model to study counterion interactions in a colloidal solution.  In this shell model, the counterions are restricted to move inside a spherical region about their host colloidal particle.  In particular, we apply Monte Carlo simulations to derive the energy and entropy contributions of the effective colloidal interaction.  Our result reveals an attractive electrostatic energy, which is overpowered by the osmotic repulsion among the counterions, as the latter can be well estimated by an ideal-gas approximation.  We also provide an optional algorithm that enables counterion mixing between the two counterion clouds even when the clouds do not overlap.  The residual mixing entropy of counterions gives a reduction in free energy that is comparable to the thermal fluctuation, suggesting a possible attractive mechanism between the colloidal particles under non-equilibrium condition.
} 

\maketitle

\section{Introduction}
The phenomenon of the so-called ``like-charge attraction'' among colloidal particles has been found for almost two decades\cite{Fraden94,Tinoco96,Grier96,Grier97,Grier03,Liu04,Tata08,Pollack09} in solutions of low-ionic strength.  The range of this attraction is in the order of the colloidal size, which is too large to be accounted by the van der Walls interaction.  Such attraction exists even when the counterions are monovalent, indicating a possible failure of the DLVO theory\cite{DLVO,Israelachvili}.  Even after large discussions\cite{Sogami84,Sogami05,Ise10,Overbeek87,Tinoco98,Tinoco02,Squires00,Levin99,Levin02,Baumgartl06}, the question is still quite under debate whether the like-charge attraction origins from artifacts of the experimental setup.  On the other hand, similar attractions have been observed in the system of polyelectrolyte as well\cite{Schmitz84,Schmitz93}, in which the phenomenon is often named as ``ordinary-extraordinary transition''.

Certainly counterions play a leading role in the discussions about the sources of the like-charge attraction.  However, it is important to verify if the attraction is attributed to their energy or entropy contribution.  In a salt-free solution, one learns from the DLVO or Debye-H\"uckel\cite{Mcquarrie} theory that the screening length of colloidal particles is comparable with the colloidal size.  This rather huge screening length provides a larger volume of residence for the counterions.  Hence these point-like counterions about their host colloid can be somewhat viewed as ideal-gas particles that are confined by the overall electric field of the system.  Prompted by this picture, we proposed a shell model, in which the counterions are restricted to move inside a spherical shell about their host colloidal particle.  This shell model facilitates our task in deducing the free energy of two colloidal particles along with their counterions.  Our model is given in details in Sec.~\ref{sec_model}, as the results of simulations are presented in Sec.~\ref{sec_energy} and Sec.~\ref{sec_free}.

The term ``mixing entropy'' often refers to the increase in entropy due to mixing of larger than two kinds of particles.  That is based on the understanding that the mixing of same kind of particles results in essentially no change of entropy in a bulk system.  However, we have shown in our previous work\cite{ms_gibbs} that there exists a non-extensive residual mixing entropy even if the mixing particles are of the same kind.  This residual entropy may be negligible in a bulk system, but in a mesoscopic system of colloidal particles, the entropy effect due to the mixing of counterions results in a free energy reduce that is comparable with the thermal energy.  In this current work, we proceed with this idea of mixing, as we find from the simulation result in Sec.~\ref{sec_mixing} that the entropy effect due to mixing of counterions lead to a possible like-charge attraction in the non-equilibrium case, where mixing is incomplete over the short-time scale.

\section{Attractive electrostatic energy in the DLVO theory}
According to the DLVO theory\cite{DLVO,Israelachvili}, the effective electrostatic interaction between a pair of colloidal particles is
\begin{equation}
  \beta F = 64 \pi R \rho_{\infty} \frac{\gamma^2}{\kappa^2} e^{-\kappa(d-2R)} \, ,
\label{eqn_DLVO}
\end{equation}
where
\begin{equation}
  \gamma \equiv \tanh \left( \frac{e \psi_0}{4 k_{\rm{B}}T} \right) \, ,
\end{equation}
and
\begin{equation}
  \kappa^2 = \frac{e^2}{\epsilon k_{\rm{B}}T} \cdot 2\rho_{\infty} 
\end{equation}
if the salt ions are monovalent.  Also $\rho_{\infty}$ is the number density of salt ions, $R$ is the radius of a colloidal particle, and $d$ is the distance between the two colloidal particles.  The electric potential at the colloidal surface $\psi_0$ is
\begin{equation}
  \psi_0 = -\frac{Ze}{4\pi R \epsilon (1+\kappa R)} 
\end{equation}
according to the Debye-H\"uckel theory ($-Ze$ is the charge of each colloidal particle).  Note that the effective energy described in Eq.~\ref{eqn_DLVO} is actually a free energy (potential of mean force), as it includes entropic contributions of small ions.  To derive the pure electrostatic energy, one can apply the thermodynamic relation $\partial (\beta F) / \partial \beta = U$.

\begin{figure}
  \resizebox{0.8\textwidth}{!}{%
    \includegraphics{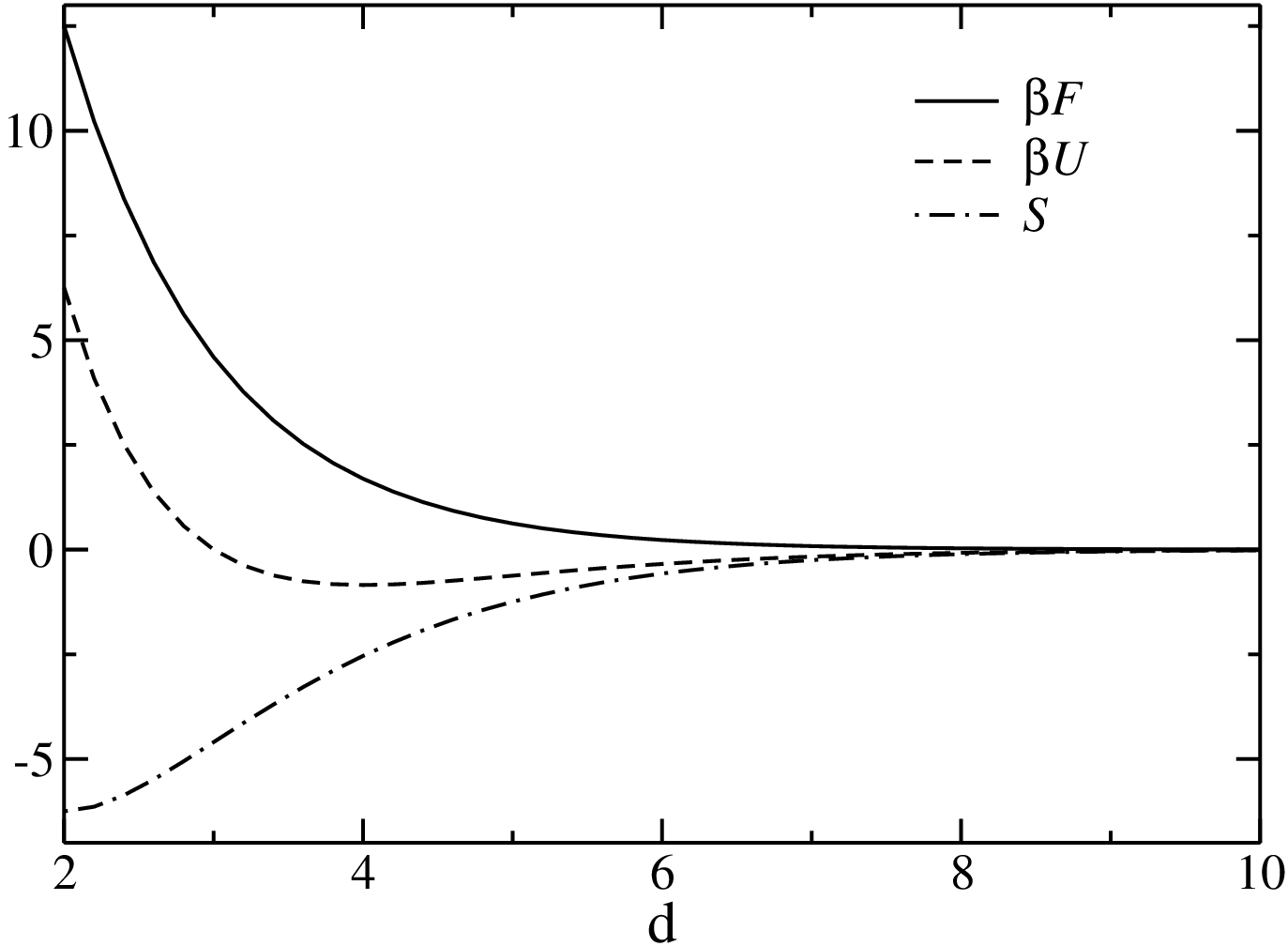}
  }
  \caption{Comparison of the free energy, energy and entropy of the DLVO theory.  In this example we use $Z=100$ and $t=100$, as the quantities are presented in dimensionless units.}
  \label{fig_DLVO}
\end{figure}

For the salt-free case, $\rho_{\infty}$ is small but not exactly equal to zero due to the dissociation of water molecules.  Such a low value of $\rho_{\infty}$ contributes to a screening length in the order of one micrometer, as we use $\kappa R =1$ in this work.  Fig.~\ref{fig_DLVO} shows the comparison of $\beta F$, $\beta U$ and $S$ for the case that $Z=100$ and $t = 100$, as we use dimensionless units such that $R=1$ and $k_{\rm{B}}=1$, and the reduced temperature $t$ is defined as the inverse of the Bjerrum length $\ell_{\rm{B}} = e^2/(4\pi \epsilon k_{\rm{B}} T)$.  The entropy is derived by $S = \beta U - \beta F$, with the reference point that $S(d \rightarrow \infty) =0$.  Note that the pure energy part $\beta U$ exhibits attractive interaction in the mid-range of colloidal separation, while such attractions is overpowered by the entropic part, which shows osmotic repulsions of the counterions.

The DLVO expression of the effective electrostatic energy, as given in Eq.~\ref{eqn_DLVO}, is based on the Derjaguin approximation\cite{Israelachvili}, which works only for the situation that the screening length is much smaller than the colloidal size.  While this approximation fails in the salt-free case, we perform Monte Carlo simulations to re-verify this feature of attraction in the pure-energy part in the following sections.  Meanwhile, we try to examine the energy and entropy contributions with more details, and check if there exists a scenario for the effective like-charge attractions between colloidal particles.

\section{Model}
\label{sec_model}
We set in our model that each colloidal particle has a net charge $-Ze$ located at its center, and the radius of a colloidal particle is $R$.  The charge of the colloidal particle is surrounded by $Z$ neutralizing counterions, each bearing a charge $+e$.  These small counterions, modeled by point-like particles with no size, are confined in a spherical region of radius $2R$ about the center of the colloidal particle, as they are also excluded from the colloidal volume.  The reason that $2R$ was chosen to be the size of the counterion cloud is mainly from of the fact that the size of the counterion cloud (and hence the colloidal screening length) in the salt-free regime is compatible with that of a micrometer-sized colloidal particle.  We then perform Monte Carlo simulations, while during the course of each simulation we fix the distance $d$ between the two colloidal particles and derive the corresponding effective colloidal interaction from the average electrostatic energy.  Note that the shell model approach has been extensively studied\cite{Chaikin84,Deserno01}.  While the original version of the cell model allows no fluctuations of counterions among different cells, we use an optional algorithm in which the counterions are allowed to transfer even if the cells are not joined together.  Through this optional algorithm, along with the simplicity of the cell model approach, our objective is to study the effects of counterion mixing in this work.

In this work the relevant energy is computed by considering the electrostatic interactions only, and we use dimensionless variables to represent relevant physical quantities.  In particular we set $R=1$. Therefore the total electrostatic energy among the colloidal particles and their counterions can be written as
\begin{equation}
  U = \sum_{\mbox{all\ charges}} \frac{z_i z_j}{r_{ij}}.
\end{equation}
where $z_i$ is the valence of the $i$th charge ($z_i=1$ for counterions and $z_i= -Z$ for colloidal particles), and $r_{ij}$ is the distance between the $i$th and $j$th charge.  Moreover, we introduce a reduced temperature $t$ which corresponds to the ratio of the colloidal radius over the Bjerrum length.  For aqueous solutions at room temperature, $\ell_{\rm{B}} \approx$ 7$\AA$, and therefore $t$ is in the order of several hundreds to a thousand (since the radius of colloidal particles is in the order of one micrometer).  The magnitude of the reduced temperature is relatively large compared with the electrostatic interaction between a single counterion and its host colloidal particle, or the interaction between two neighboring counterions.  For most of the time, the latter is small compared with the average thermal energy in our system with sparsely distributed counterions.  As a consequence, the pairwise electrostatic interactions are rather weak, as the force exerted on each counterion are contributed from the collective charge distribution.

In our current model we neglect the ``image charge'' effect\cite{Messina02,Coalson94} caused by the difference in dielectric constants between the colloidal particle and surrounding aqueous solution.  It has been found that such an image charge effect is crucial in colloidal solutions with multivalent counterions\cite{Messina02}. Although the influence is important as well for systems with monovalent counterions\cite{Coalson94}, we have not considered in the present work as the emphasis of the present work is to bring out the importance of mixing effect.

To derive effective potentials and correlations, we perform Monte Carlo simulations via the Metropolis algorithm.  We choose $Z=100$ and $Z=1000$ in our simulations, and for each simulation we assign a random configuration for the initial counterion distribution, in which the positions of counterions obey a uniform distribution. The initialization is then followed by $10^9$ simulation updates.  The same procedure is repeated over 10 times to derive average physical quantities. For the cases where the counterion clouds do not overlap, we provide an optional mixing algorithm which allows counterions to jump between the counterion clouds, as this algorithm enables number fluctuations of counterions in a cloud.

\section{Energetic contribution of the electrostatic interaction}
\label{sec_energy}
\begin{figure}
  \resizebox{0.8\textwidth}{!}{%
    \includegraphics{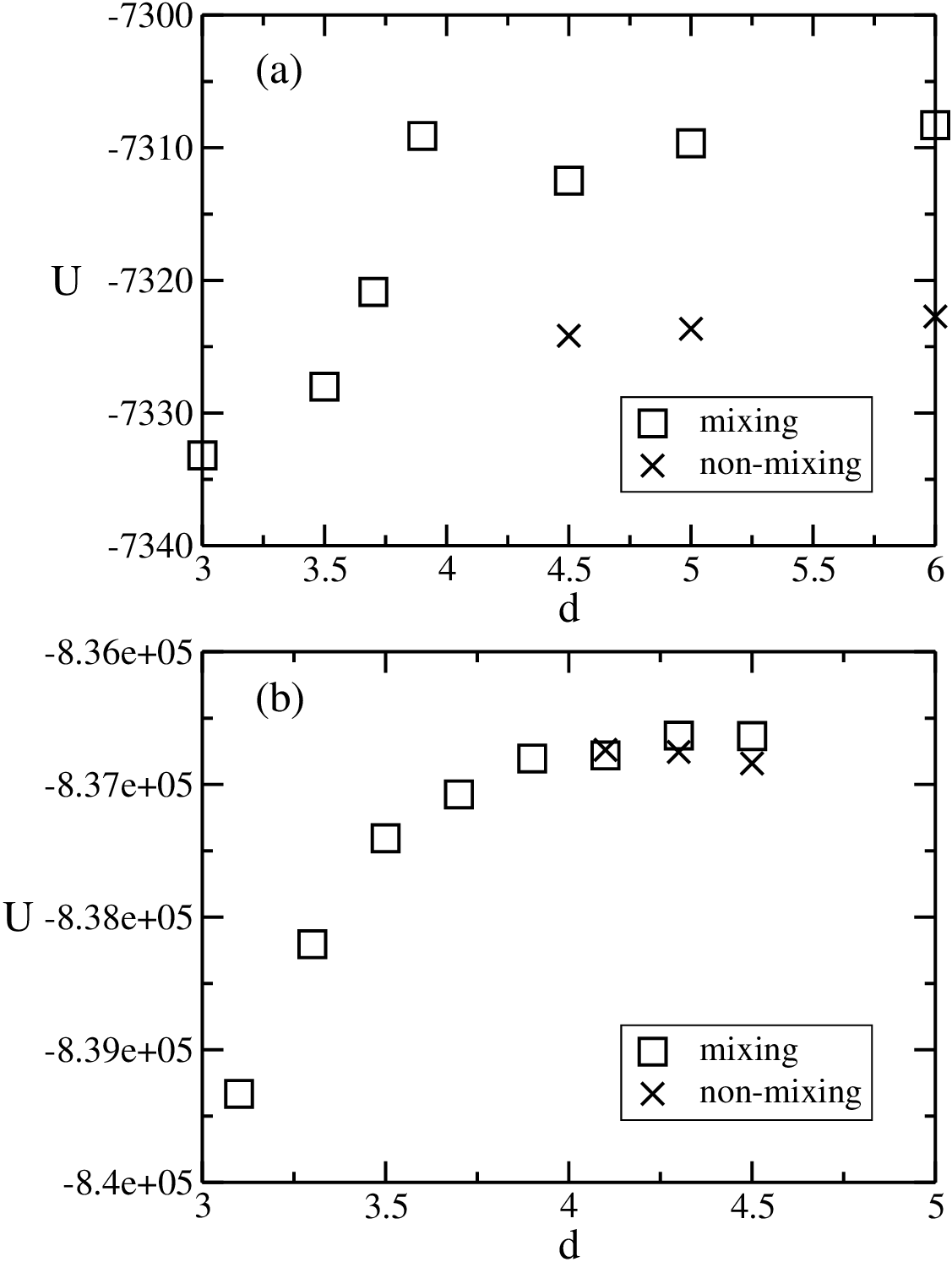}
  }
  \caption{Electrostatic energy versus the colloidal distance of our shell model for (a) $Z=100$ and (b) $Z=1000$.  The average energy is derived from Monte Carlo simulations with $t=100$.  Squares represent the results where mixing of counterions over the two clouds is enabled, while crosses represent the results of the non-mixing case.}
  \label{fig_U_d}
\end{figure}

The result of average electrostatic energy is shown in Fig.~\ref{fig_U_d} for the case $Z=100$ and $t=100$.  From the squared data (the ``non-mixing'' case will be studied in a later section) we observe attractive electrostatic interactions as the counterion cloud starts to overlap.  Within the range of our observation the magnitude of attractive energy is smaller than the thermal energy for the case $Z=100$, while for $Z=1000$ the attractive energy becomes significantly larger.  This is in agreement with the DLVO theory, although the latter is based on the Derjaguin approximation\cite{DLVO}.  In the next sections we shall obtain the free energy of the system and find that the attractive part in electrostatic energy when the counterion clouds are overlapping will be overshadowed by the entropic (osmotic) repulsions.

The source of the attractive electrostatic energy can be understood using the following argument:  since the distribution of counterions is approximately spherical, the net force between colloids approaches zero as the counterion clouds starts to separate.  While the clouds overlap, the counterions inside the overlapped region are less screened and therefore can feel attractions from the colloidal cores.  However, this attractive energy is reduced by the excessive electrostatic repulsions among counterions in the overlapped region.

\begin{figure}
  \resizebox{0.8\textwidth}{!}{%
    \includegraphics{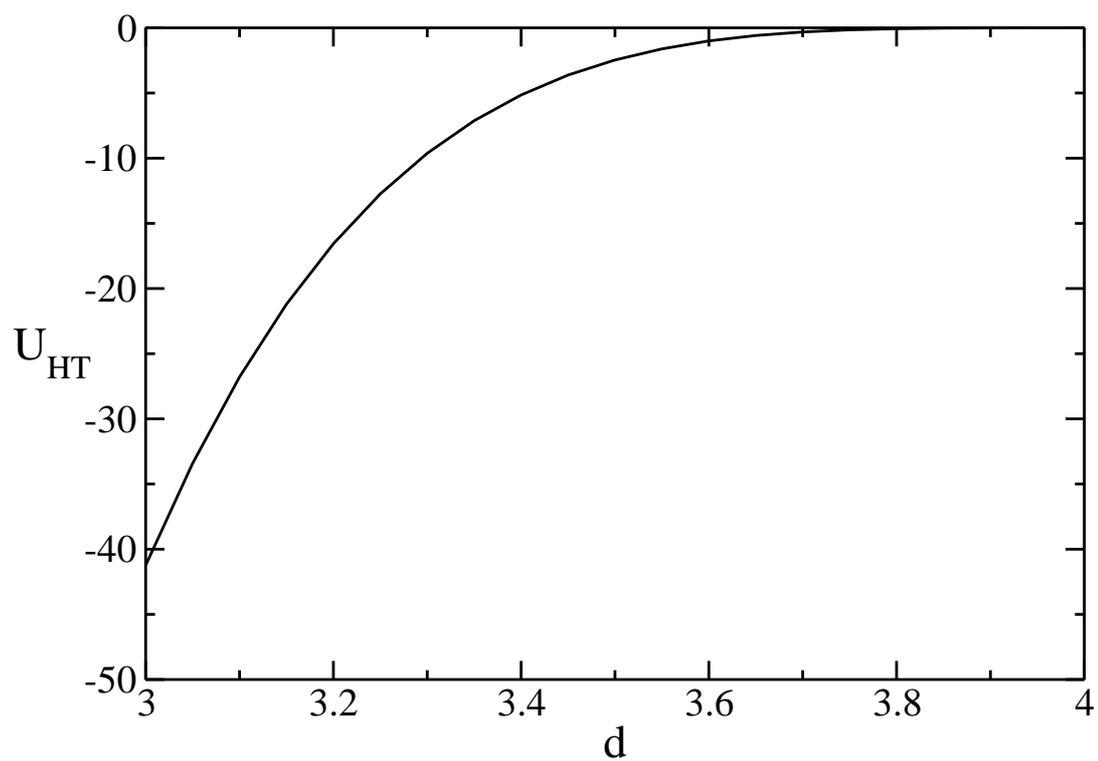}
  }
  \caption{A mean-field approximation of the electrostatic energy.  The result is derived assuming the counterion distribution about each colloid is uniform.}
  \label{fig_U_HT}
\end{figure}

To illustrate this attractive mechanism, we make a simple calculation by assuming that the counterion charges about each host colloid are uniformly distributed.  By summing over the Coulomb energy between all charges over the two colloids (except for the self energy within each colloid), one can derive
\begin{equation}
  U_{HT} \equiv U(T\rightarrow \infty) = - \frac{\pi^2 \rho^2}{90d} \cdot (2400 R^6 - 6144 R^5 d + 640 R^3 d^3 - 120 R^2 d^4 + d^6) -\frac{Z^2}{d}\, ,
\end{equation}
where $\rho \equiv Z/(7 \cdot 4 \pi R^3/3)$ is the homogeneous counterion density.  The behavior of $U_{HT}$ is shown in Fig.~\ref{fig_U_HT}, as it reveals an attractive energy of similar magnitude compared with our simulation results.

\begin{figure}
  \resizebox{0.8\textwidth}{!}{%
    \includegraphics{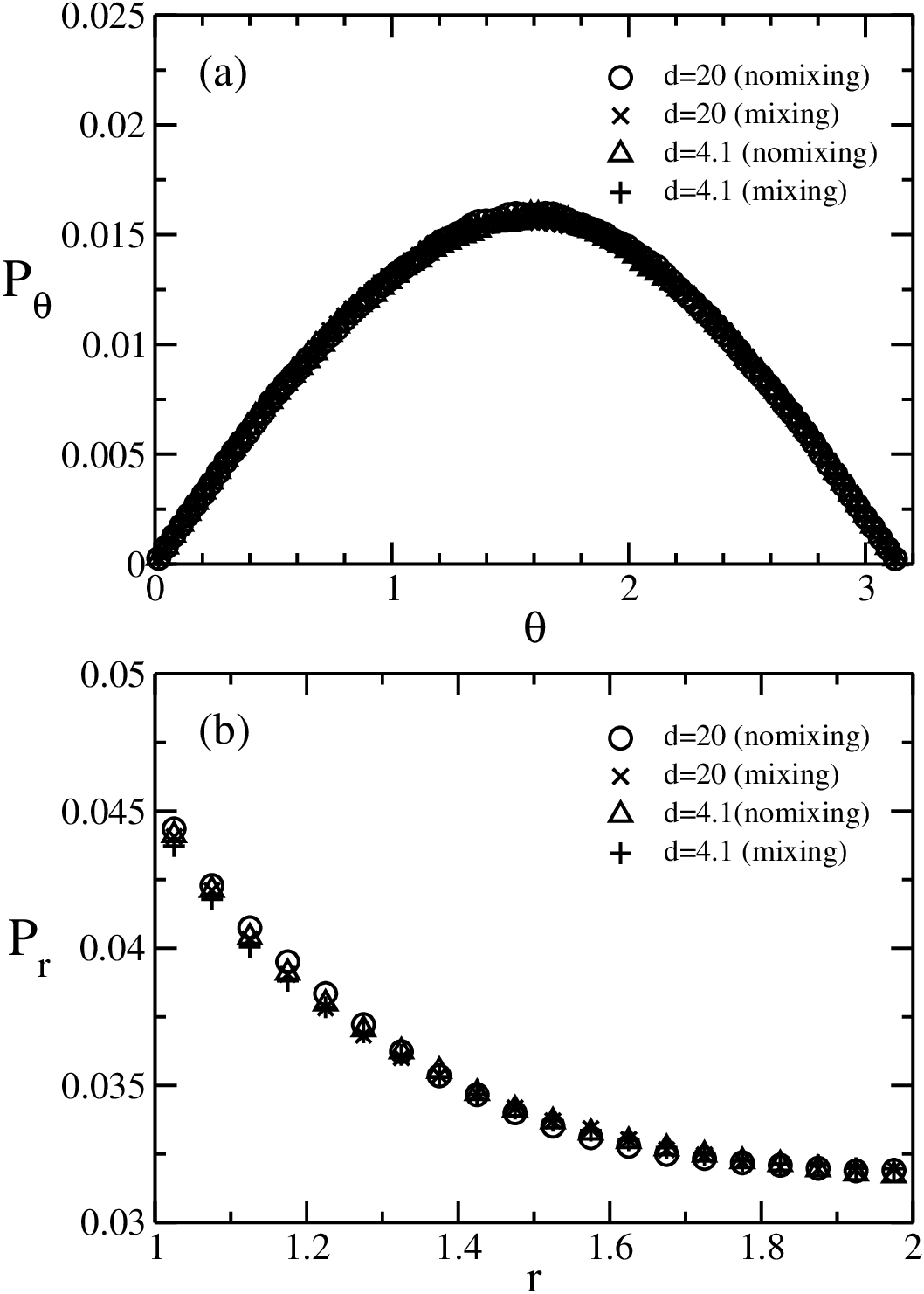}
  }
  \caption{Counterion distributions in the non-overlapping case.  (a) The angular distribution. (b) The radial distribution.}
  \label{fig_P_theta_r}
\end{figure}

Meanwhile, in our simulations we have checked about the counterion distribution for both the non-overlapping and overlapping cases.  The radial and angular distributions of counterions in the non-overlapping case are given in Fig.~\ref{fig_P_theta_r}.  The angle $\theta$ represents the azimuthal angle with respect to the line joining the centers of colloidal particles, as one finds that the counterions for the non-overlapping case are distributed uniformly angular-wise, as their distribution about the azimuthal angle is sinusoidal.  Moreover, the radial distribution exhibits a monotonic decay, although the distribution is not identical to that of the DLVO result, as the decay is milder due to our sharp cutoff at the shell boundary.

\begin{figure}
  \resizebox{0.8\textwidth}{!}{%
    \includegraphics{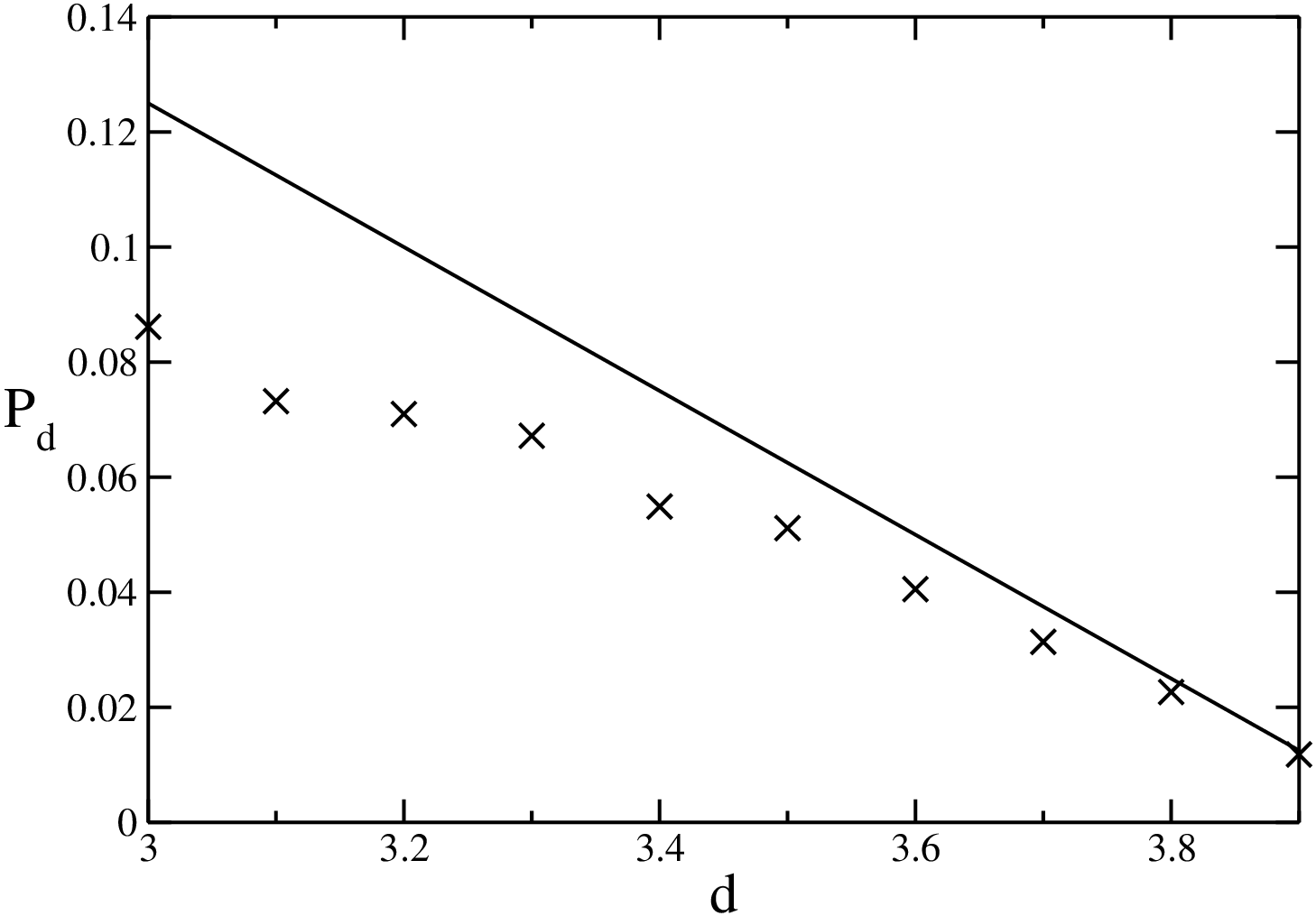}
  }
  \caption{The number fraction of counterions within the azimuthal angle in the overlapping region for various $d$.  The line corresponds to the case of a uniform angular distribution in each counterion cloud.}
  \label{fig_P_d}
\end{figure}

For the case that counterion clouds are overlapping, the result of counterion angular distribution is represented by a plot of counterion fraction compared with that of a uniform distribution within the azimuthal angle in the overlapping region.  From Fig.~\ref{fig_P_d} one finds that when the counterion clouds are overlapping, some of the counterions diffuse out from the overlapping region, as the fraction becomes smaller compared with that for the non-overlapping case, the latter possessing a uniform angular distribution.  The drop in number of counterions may result from osmotic or electrostatic repulsions of the counterions (or both) in the overlapping region.

\begin{figure}
  \resizebox{0.8\textwidth}{!}{%
    \includegraphics{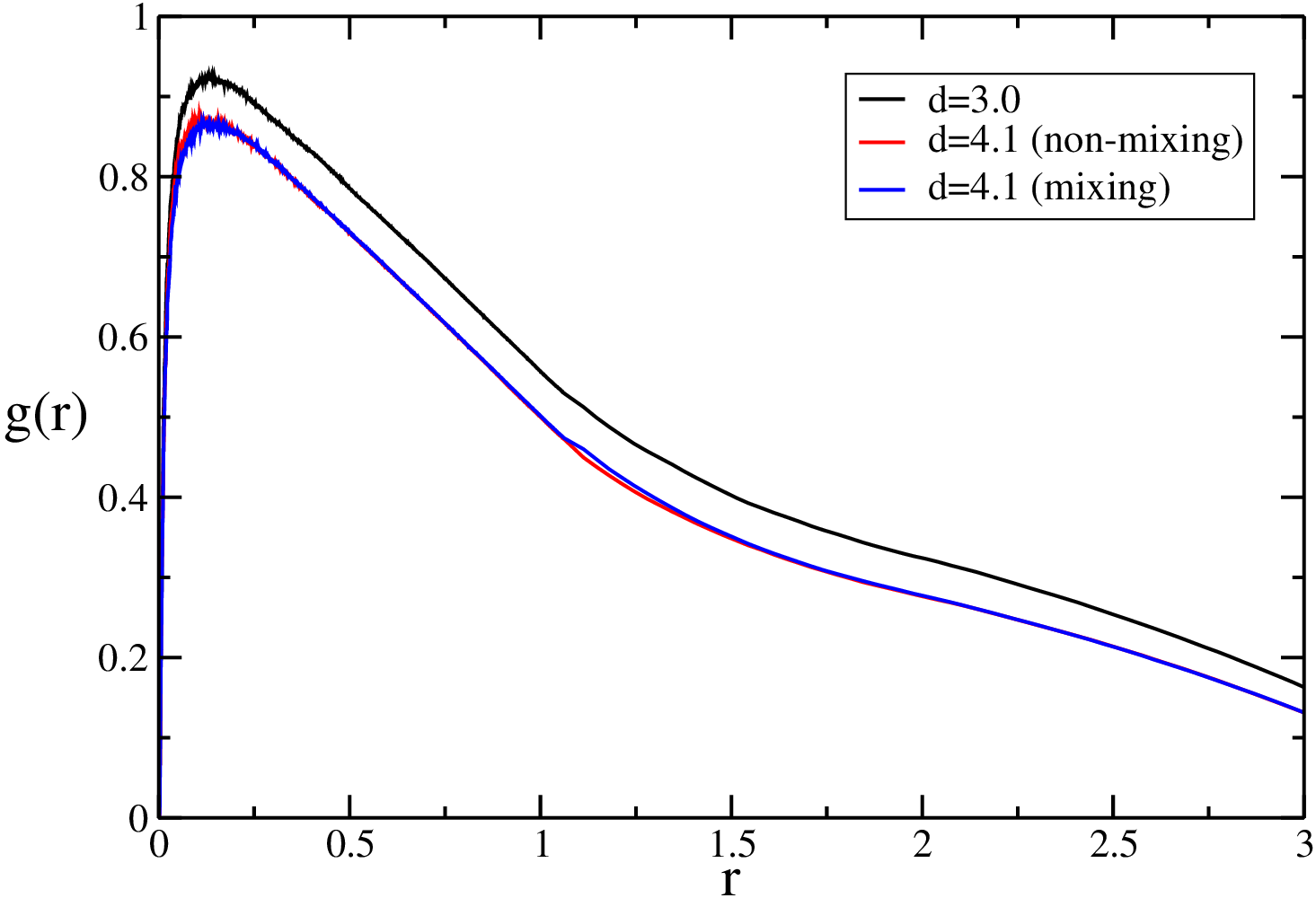}
  }
  \caption{Counterion-counterion pair distribution function $g(r)$ for $d=3.0$ and $d=4.1$ (including both the non-mixing and mixing cases).}
  \label{fig_pair_dist}
\end{figure}

The counterion-counterion correlation function is obtained and plotted for the non-overlapping ($d=4.1$) and overlapping ($d=3.0$) cases in Fig.~\ref{fig_pair_dist}.  While there seems to be negligible difference between the mixing and non-mixing cases, the correlation function shows a higher peak for the overlapping case compared with that for the non-overlapping case.  From such a higher peak one may speculate that the counterions tend to 'aggregate' in the overlapping region, although there are chances that the result is due to the difference in geometric constraints between the non-overlapping and overlapping cases.

To verify whether there exists any attraction in the overall effective energy, we continue to derive the free energy in the following paragraphs, as the result shows that the electrostatic attraction described here is indeed overpowered by the entropy effect, although the result shows the existence of another possible mechanism of like-charge attraction due to counterion mixing.

\section{Derivation of the free energy}
\label{sec_free}
\subsection{Entropy in the high-temperature limit}
\label{entropy_HT}
\begin{figure}
  \resizebox{0.8\textwidth}{!}{%
    \includegraphics{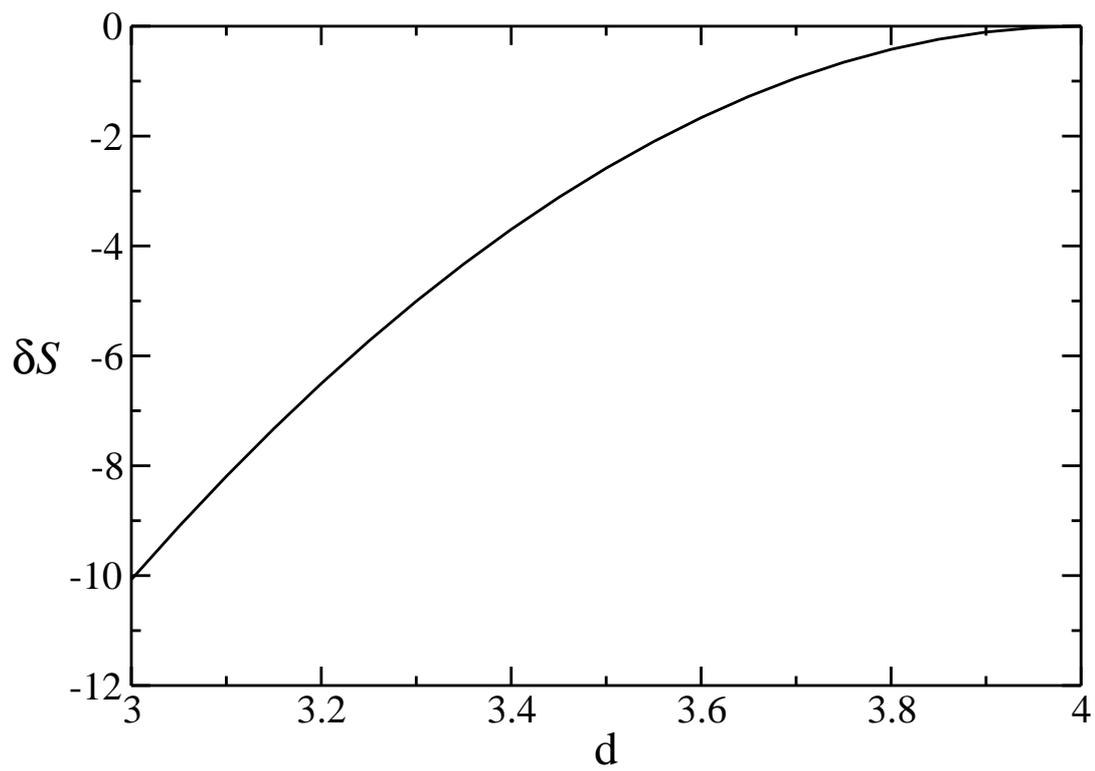}
  }
  \caption{Change of counterion entropy in the high-temperature limit due to volume decrease.  In our shell model, the residence volume of counterions decreases once the two counterion clouds overlap, resulting in an entropy decrease.  The result is derived using an ideal-gas law of entropy.}
  \label{fig_entropy}
\end{figure}

To derive the free energy one first needs the value of $\beta F$ when $\beta \rightarrow 0$, which is equal to the negative entropy in the high-temperature limit.  In our model the counterions become ideal in the high-temperature limit except for the restriction in their occupying volume.  As the colloidal particles approach, the counterion clouds overlap and the volume of occupation starts to reduce.  With the use of the ideal entropy (recalling that $k_{\rm{B}}=1$ in our reduced unit)
\begin{equation}
  \left. S \right| _{\beta=0} = 2N \ln V \, ,
\end{equation}
one can derive the change in entropy due to volume decrease:
\begin{eqnarray}
  \left. \delta S \right| _{\beta=0} &=& 2N \ln \left( \frac{V-\Delta V}{V} \right) \nonumber \\
   &=& -2N \ln \left( \frac{224}{96+48d-d^3} \right) 
\end{eqnarray}
for $2<d<4$, where $d$ is the distance between the centers of the two colloidal particles (recalling that $R=1$ in our model).  From Fig.~\ref{fig_entropy} one finds that the entropy decreases monotonically as the distance becomes smaller.  For example, $\delta S = -2.5836$ when $d=3.5$, which provides an increase in free energy that is crucial compared with thermal fluctuation.  As we shall see next, this entropy decrease in the high-temperature limit is dominating over the electrostatic attraction, and it will be corrected slightly by considering the free energy change towards the finite-temperature regime.

\subsection{Method}
\label{free_method}
In this subsection we propose a new method to derive the Helmholtz free energy via Monte Carlo simulations.  Commonly one uses the relation
\begin{equation}
  \left. \frac{\partial \beta F}{\partial \beta} \right|_{\mbox{volume fixed}} = U\, ,
\end{equation}
and derive the free energy by integrating the internal energy over the inverse temperature $\beta \equiv 1/t$.

We start by rewriting the partition function at inverse temperature $\beta_0$
\begin{eqnarray}
  Q_{\beta_0} &=& \sum_\nu e^{-\beta_0 U_\nu} \nonumber \\
            &=& \sum_\nu e^{-\Delta \beta U_\nu} e^{-\beta_1 U_\nu} \nonumber \\
            &=& \left \langle e^{-\Delta \beta U} \right \rangle_{\beta_1} \cdot Q_{\beta_1} \, ,
\label{partition}
\end{eqnarray}
where $\Delta \beta = \beta_0 -\beta_1$ and $\langle \rangle_{\beta_1}$ implies the thermodynamic average at the inverse temperature $\beta_1$. From Eq.~\ref{partition} one can derive the free energy difference between $\beta_0$ and $\beta_1$:
\begin{equation}
  (\beta F)_{\beta_0} - (\beta F)_{\beta_1} = - \ln \left( \frac{Q_{\beta_0}}{Q_{\beta_1}} \right) = -\ln \left \langle e^{-\Delta \beta U} \right \rangle_{\beta_1} \, . 
\label{free_diff}
\end{equation}
Through the average in Eq.~\ref{free_diff} one gains the difference in $\beta F$ over one simulation. This relation is exact and one does not have to apply any other integrating approximation. In practice, we choose $\beta_1 = \beta_0 /M$, in which $M$ is an integer (eg. $M=4$ is used in our simulations). As $M$ is not very large, the importance sampling still works and one can gain the value $\beta F$ by applying Eq.~\ref{free_diff} iteratively. For a system in which the dynamics may be hampered by various local minima, this technique can be applied along with the method of replica exchange\cite{Swendsen86}, so that free energy is derived while frustrated dynamics can be reduced during the course of parallel simulations over various temperatures.

\subsection{Result}
In this work we apply the above method to derive $\Delta \beta F$ between $t=100$ and $t=25600$, as the results for $d=3.5$ and $d=20$ are shown in Table~\ref{table}. Note that the residual difference between $t=25600$ and $t \rightarrow \infty$ is approximated by $(\beta U)_{t=25600}$.  To compare the overall free energy, one has to count additionally the free energy difference in the high-temperature limit as derived in Sec.~\ref{entropy_HT}.  For example, from Fig.~\ref{fig_entropy} we find this entropy difference in the high-temperature limit to be -2.5836 between $d=3.5$ and $d=20$.  Along with the result (of the mixing case) in Table~\ref{table}, we conclude that the overall free energy ($\beta F$) of $d=3.5$ is higher than that of $d=20$ by the amount 2.2805.  Therefore the electrostatic attractive energy observed in Sec.~\ref{sec_energy} is heavily outweighed by the entropy effect, the latter of which is mostly attributed to the entropy of counterions in the high-temperature (ideal-gas) limit.

\section{Effect of mixing}
\label{sec_mixing}
In this work we provide an optional mixing algorithm for cases that the two counterion clouds do not overlap.  As to a real colloidal solution, counterions are allowed to diffuse between clouds, thus enabling counterion number fluctuation in each cloud.  Nevertheless, such diffusion takes a relatively longer time, and mixing between two counterion clouds can be rather incomplete during the observation time in a typical experiment.  To estimate this effect of mixing, we turn our attention to the results for the mixing and non-mixing cases in our simulations.

From Fig.~\ref{fig_U_d}, one finds that the electrostatic energy of the non-mixing case is lower than that of the mixing case.  However, from the second law of thermodynamics\cite{Chandler}, one knows that the free energy in the mixing case must be lower than that in the non-mixing case, because there is less restriction in the mixing scenario.

To investigate this free energy difference, we look up in the data in Table~\ref{table} for the case of non-overlapping clouds $d=20$.  The result shows that $\Delta \beta F$ of the non-mixing case is smaller than that of the mixing case by the amount 0.2259.  This deficit is outweighed by the entropy difference between the mixing and non-mixing cases in the high-temperature (ideal gas) limit\cite{ms_gibbs}.  A simple calculation using the ideal gas entropy $S=N \ln  V$ leads to a mixing entropy
\begin{equation}
  S_{\rm{mixing}} = \ln \left[ \frac{2^{2Z}(Z!)^2}{(2Z)!}  \right] \, .
\label{gibbs_entropy}
\end{equation}
For the case $Z=100$ one obtains $S_{\rm{mixing}}= 2.8762$.  Therefore $\beta F$ of the mixing case is ultimately smaller than that of the non-mixing case by the amount 2.6503, which shows that the free energy difference is quite compatible with the thermal energy in its magnitude.  For the case $Z=1000$, we obtain that the free energy difference between the mixing and non-mixing cases at $d=20$ is 6.1876 using the same analysis.

When the thermodynamic equilibrium is achieved, mixing of counterions is complete for all colloidal distances.  However, the mixing between counterions of both clouds may be rather incomplete within the limited time of observation at a large colloidal separation.  To encourage the mixing process, colloidal particles will be dragged towards each other in the short-time interval to reduce the total free energy.  Meanwhile, there exists a balancing mechanism in which the counterion clouds tend to stay less overlapped to avoid osmotic repulsion.  Hence our result gives a hint that there exists a non-equilibrium mechanism of like-charge attraction that drives the colloidal particles towards some optimized separation, and this non-equilibrium mechanism is attributed to the residual mixing entropy of counterions.

To follow the ``mixing algorithm'' discussed in this manuscript, if one considers the 'real' situation, where there are residual salt ions or dissociated water ions in the neighborhood of the counterion clouds, the counterion fluctuations of a single cloud can be easily facilitated through exchange with its neighboring solution, provided that there are ample amount of residual ions.  Probably this explains why an effective like-charge attraction cannot be observed in real experiments for the case $Z=100$, because the counterion density within the cloud is not much deviated with that in the ambiance.  On the other hand, for the case $Z=1000$ or even $Z=10000$, the ionic concentration within the cloud is quite large compared with its neighborhood.  For this situation there exist less counterion fluctuations if the clouds do not stay close, since mixing cannot be easily achieved by the exchange of counterions of a single cloud with its surrounding, where residual ions are relatively sparsely distributed.

The role of counterion fluctuations and its link to like-charge attractions has first been studied in the Sogami-Ise theory\cite{Sogami84}, which is basically an extension of the DLVO theory and thus a mean-field theory, in which counterion fluctuations are not fully considered. Based on the DLVO theory they obtained an effective attraction among colloidal particles via a derivation about the Gibbs free energy instead of the Helmholtz free energy.  While the counterion fluctuation considered in the Sogami-Ise theory is limited, due to its mean-field nature, counterion fluctuations are better described in the work of Carbajal-Tinoco {\it et al.}\cite{Tinoco02}.  While in the above work the authors attributed the like-charge attraction to charge inversions about the colloids due to counterion fluctuations, we have not observed in our simulations any like-charge attraction from the viewpoint of pure electrostatic interactions, at least in the current shell model description.  On the other hand,  the result of our work shows that like-charge attraction may arise due to entropic difference between the mixing and non-mixing cases, as we observe a mixing entropy even for identical particles (i.e, counterions).

In the work of Polin {\it et al.}\cite{Grier07}, they proposed a phenomenological expression of effective colloidal interaction that gives both repulsive and attractive parts.  The attractive part has a longer range than the repulsive part, suggesting that effective attractions emerge beyond the range where colloidal charges are fully screened.  While the authors speculated that the long-range attraction is due to nonmonotonic distribution of counterions about their their host colloid, we have not derived such a distribution in our current simulation studies.  Nevertheless, the comparison between the mixing and non-mixing results also indicates a possible like-charge attraction whose range is longer than the diameter of the counterion shell, the latter being comparable to the screening length.  This is in agreement with the phenomenological expression by Polin {\it et al.}\cite{Grier07}.

\section{Conclusion}
\label{Conclusion}
In summary, we apply our shell model to study the effects of counterionic energy and entropy contributions to the effective free energy between a pair of colloidal particles in the salt-free regime.  We observe that the electrostatic energy shows an attractive part at shorter colloidal distances, although this is overpowered by the osmotic repulsion between counterions, the latter resulting from the decrease of their residence volume.  As our model assumes that counterions about each colloidal particle must reside in spherical shell, such a restriction may lead to a drop in volume that may be over estimated as the counterion clouds overlap.  We expect the volume decrease and thus the entropy decrease to be smaller if the volume of counterion residence is allowed to deform in shape.  Therefore we still cannot rule out the possibility that the electrostatic attraction might eventually dominate over the osmotic repulsion.

Meanwhile, the residual mixing entropy between the two counterion clouds as introduced in our previous work\cite{ms_gibbs} is further investigated in our current study.  We argue that this residual entropy can lead to an effective attraction between colloidal particles in the short-time interval, where mixing of counterions over the two colloids is still incomplete at larger colloidal separations.  To further check about this idea, we suggest to study the hydrodynamic modes of the colloidal system using the Poisson-Boltzmann approach in a future work.  Finally, the argument about the residual mixing entropy is simple and general enough that it may be applied to explain like-charge attractions observed in similar systems such as polyelectrolyte\cite{Schmitz84,Schmitz93}.

This work was supported by the National Science Council of the Republic of China under Grant No. NSC-100-2112-M-008-005.  The author would like to thank for support from NCTS under focus group Life and Complexity.

\clearpage
\begin{table}
  \begin{tabular}{|c|c|c|c|}
   \hline
   $t_1-t_2$ & $\Delta \beta F (d=3.5)$ & $\Delta \beta F (d=20)$ (mixing) & $\Delta \beta F (d=20)$ (non-mixing) \\
   \hline
   $100-400$ & -54.5479 & -54.3311 & -54.5000 \\
   \hline
   $400-1600$ & -13.4914 & -13.4268 & -13.4674 \\
   \hline
   $1600-6400$ & -3.3629 & -3.3462 & -3.3591 \\
   \hline
   $6400-25600$ & -0.8403 & -0.8366 & -0.8392 \\
   \hline
   $100-\infty$ & -72.5226 & -72.2195 & -72.4454 \\
   \hline
  \end{tabular}
  \caption{Free energy difference $\Delta \beta F$ between temperatures $t_1$ and $t_2$. These free energy differences are derived based on the algorithm as described in Sec.~\ref{free_method}. For the cases where the counterion clouds do not overlap (i.e., $d>4$) in our model, we provide an optional simulation algorithm in which counterions are allowed to jump between clouds. This jumping algorithm enables mixing and therefore counterion number fluctuation. For $d=20$ we list out the results of both algorithms.}
  \label{table}
\end{table}

\clearpage

\clearpage

\clearpage

\clearpage

\clearpage

\clearpage

\clearpage

\clearpage

\end{document}